\documentclass[sigconf,10pt,nonacm]{acmart}

\setcopyright{none}

\renewcommand\footnotetextcopyrightpermission[1]{} % removes footnote with conference information in first column

\usepackage{graphicx} % Required for inserting images

\usepackage{tikz}
\usepackage{amsmath}

\usepackage{filecontents}

\usepackage{enumitem}
\usepackage{xspace}
\usepackage[frozencache=true, cachedir=minted-cache]{minted}
\usepackage{cleveref}
\usepackage{siunitx}
\usepackage{booktabs}
\usepackage[utf8]{inputenc}
\usepackage[T1]{fontenc}

\newcommand{\sys}{NaSh\xspace}

\newcommand{\sx}[1]{(\S\ref{#1})}

\newcommand{\heading}[1]{\vspace{1pt}\noindent\textbf{#1.}\enspace}
\newcommand{\ttt}[1]{\mintinline[fontsize=\normalsize, breaklines, breakafter=-]{js}{#1}}

\newcommand{\figref}[1]{Fig.~\ref{#1}}
\newcommand{\secref}[1]{\S{}\ref{#1}}

\usepackage{caption}
\begin{document}

\title{\sys{}: Guardrails for an LLM-Powered Natural Language Shell}

\author{Bimal Raj Gyawali}
\affiliation{
  \institution{Unaffiliated}
  \country{}
}
\email{gyawali.rajbimal35@gmail.com}

\author{Saikrishna Achalla}
\affiliation{
  \institution{University of California, Los Angeles}
  \country{}
}
\email{saiachalla@cs.ucla.edu}

\author{Konstantinos Kallas}
\affiliation{
  \institution{University of California, Los Angeles}
  \country{}
}
\email{kkallas@ucla.edu}

\author{Sam Kumar}
\affiliation{
  \institution{University of California, Los Angeles}
  \country{}
}
\email{samkumar@cs.ucla.edu}

%% Abstract
%% Note: \begin{abstract}...\end{abstract} environment must come
%% before \maketitle command
\begin{abstract}
We explore how a shell that uses an LLM to accept \emph{natural language input} might be designed differently from the shells of today.
As LLMs may produce unintended or unexplainable outputs, we argue that a natural language shell should provide guardrails that empower users to recover from such errors.
We concretize some ideas for doing so by designing a new shell called \sys{}, identify remaining open problems in this space, and discuss research directions to address them.

\end{abstract}

\maketitle

\section{Introduction}

LLMs are evolving from chatbots into agents capable of performing tasks on behalf of users~\cite{li2024llmagent, lewis2021rag, varshney2023llmagent, martineau2024llmagent}.
Anthropic recently demonstrated an LLM agent capable of interacting with computers via keyboard and mouse inputs~\cite{anthropiccomputeruse}, and researchers have fine-tuned LLMs for the specific task of outputting actions (e.g., API calls)~\cite{patil2024gorilla}.
Just as the transition from classical shells (e.g., Bourne shell~\cite{bourne1978shell}) to graphical shells (e.g., Windows Explorer~\cite{sullivan1996windows95ui}) made computers accessible to non-programmers~\cite{nielsen1993noncommand, vandam1997postwimp}, LLM-based agents could make computing even more accessible.
Specifically, they could enable \emph{natural language shells} that allow users to simply ``ask'' the computer to perform a task, via text or speech.

In this paper, we ask the question:
\textbf{How might an LLM-powered natural language shell be architected differently from a classical text-based shell or a graphical shell?}
In our view, architectural differences stem from the fact that \emph{natural language interfaces and LLMs are unreliable}.
Natural language commands are ambiguous compared to code or formal specifications.
And even for an unambiguous prompt, LLMs can take unintended actions---returning wrong results to the user, affect the surrounding system in incorrect ways, or even corrupt or delete user data.
Researchers are making progress to improve the reliability of LLMs, but given their statistical data-driven nature, achieving total confidence in their outputs may be impossible.

The thesis of this paper is that \emph{an LLM-powered shell should provide system support and guardrails that protect users from unexpected 
actions taken by LLM agents and empower users to recover from them}.
Doing so, however, is complicated by the fact that shell use is significantly different from conventional software development.
Specifically, shells are often used to manipulate program-external state, such as the file system, which can be hard to reason about and test.
This forces shell use to happen as an iterative process: \textbf{Develop}, \textbf{Run}, \textbf{Inspect}, and, if necessary, \textbf{Revert} and try again.
The guardrails we develop must work with and enhance this process.

We begin by describing the iterative process of shell use and the opportunities and challenges of applying LLMs to it (\secref{sec:shell_use}).
We then sketch the design of \sys{}, a system that provides guardrails addressing these challenges  (\secref{sec:nash}).
\sys{} focuses on shell use that is limited to \emph{local single-system} state.
Finally, we discuss open problems related to natural language shells and how to extend \sys{}'s architecture to address them (\secref{sec:discussion}).
This includes how we might extend \sys{} to provide guardrails for \emph{external} API calls.

\section{Shell Use and LLMs}
\label{sec:shell_use}

\begin{figure}
    \centering
    \includegraphics[width=0.9\columnwidth]{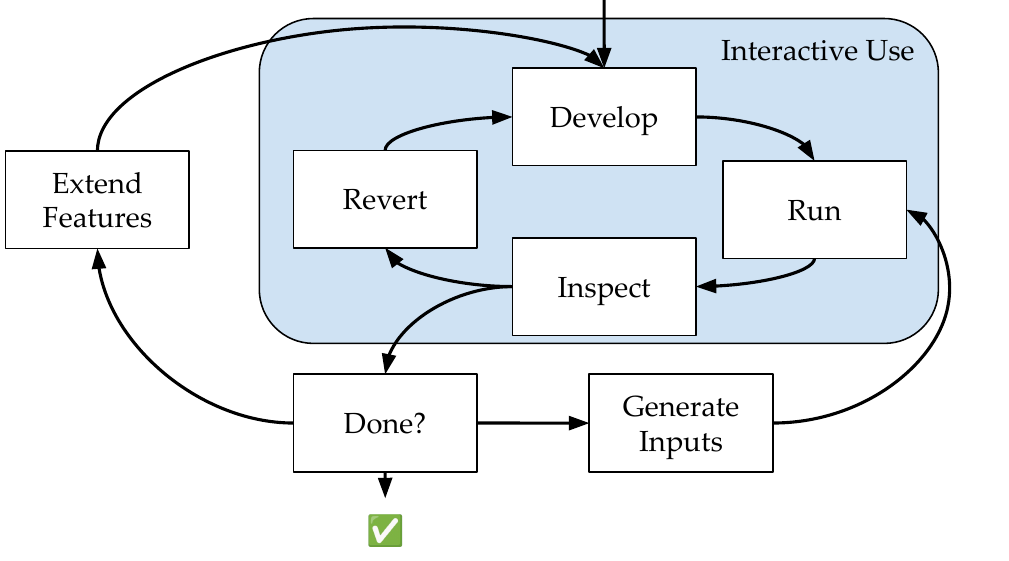}
    \caption{Shell use: inner cycle corresponds to interactive use and the complete cycle corresponds to script development.}
    \label{fig:script-development-cycle}
\end{figure}

Shells development is an iterative process (\figref{fig:script-development-cycle}), both during \emph{interactive use} and during the more extended \emph{script development}.
This process involves challenges for developers.
LLMs promise to assist with some of them,
particularly for the \textbf{Develop} phase, but exacerbate the rest.

\subsection{Interactive Commands}

Users interact with shells to perform \emph{one-off} actions, such as exploring the file system or system administration tasks.
Interactive shell usage is a cycle involving four steps:
users \textbf{Develop} the commands to execute,
they \textbf{Run} these commands and \textbf{Inspect} the system to see if they had the desired effect,
and if not, they \textbf{Revert}
to the original state before command execution.
This cycle has multiple challenges.
First, in the \textbf{Develop} step, users must learn to use both external tools and the shell language itself, extensively reading documentation and experimenting with uncommonly used commands,
e.g., when using \texttt{openssl} to generate a web certificate for testing.
Second, if the command did not work correctly, the \textbf{Revert} step requires detecting and undoing arbitrary side-effects, possibly by invoking additional complex commands.
If the state is not fully reverted after every cycle, partial results could accumulate leading to inconsistent states.

\subsection{Script Development}
\label{s:scripting}

Users write shell scripts to perform \emph{reusable} actions.
The \textsc{Unix} philosophy is for programs to be small, general components, and for shell scripts to compose these programs into meta-programs~\cite{mcilroy1978foreword}.
For example, users may create shell scripts to automate user-specific or app-specific workflows like installing a program, data processing (e.g., with \texttt{sed} or \texttt{awk}), or automatic backup (e.g., scheduled with \texttt{cron}).

Script development is an iterative process that generalizes interactive shell usage
(Fig.~\ref{fig:script-development-cycle}). 
A user writes a first version of their workflow, usually only a fragment of the whole task or targeting a subset of their data.
They then execute this version and inspect its results.
If it has not produced a correct result, the developer enters the interactive use cycles.
If the script worked as intended, then the developer goes back to the start, extending their script to implement a bigger part of their task or run on a bigger fragment of the data.

Script development introduces a key challenge compared to interactive shell usage: The user must determine that their script will behave correctly in future uses, when the inputs and environment might differ from the current one.
Combined with the lack of a precise specification, users are left without any support, having to manually determine and construct testing inputs and environments to test their scripts.

\subsection{Opportunities and Challenges of LLMs}
\label{s:llm_challenges}

LLMs
facilitate the \textbf{Develop} stage
by supporting natural language prompts: users no longer need to learn external tools and their options, or the intricacies of the shell language.

However, LLMs can produce incorrect answers or take incorrect actions.
While users of traditional shells may also make mistakes, LLMs exacerbate the problem because the user is more distant from the precise logic behind the execution, making it harder to predict what might go wrong and ultimately notice the mistakes.
For example, if an LLM-powered shell script unexpectedly produces files in a directory far from the current working directory, the user may not notice until long after the fact, whereas a user of a regular shell code may find it in the process of debugging and tracing their code.
Thus, while LLMs facilitate the \textbf{Develop} stage, they make the \textbf{Inspect} and \textbf{Revert} stages more difficult.

Another challenge is that even if the LLM is accurate, a natural language interface encourages ambiguities in the prompt.
For example, suppose that a user's prompt is, ``Delete all log files in the \texttt{/logs} directory.''
This may initially seem straightforward.
But on further thought, should files in subdirectories of \texttt{/logs} also be deleted?
How should the shell determine if a file in \texttt{/logs} is a ``log file''?
What about hidden files like \texttt{.system.log}?
While ordinary software development might encourage users to think through these issues, prompting hides them and forces users to rely on the \textbf{Inspect} and \textbf{Revert} steps to resolve such ambiguities iteratively.

\begin{figure}
    \centering
    \includegraphics[width=\columnwidth]{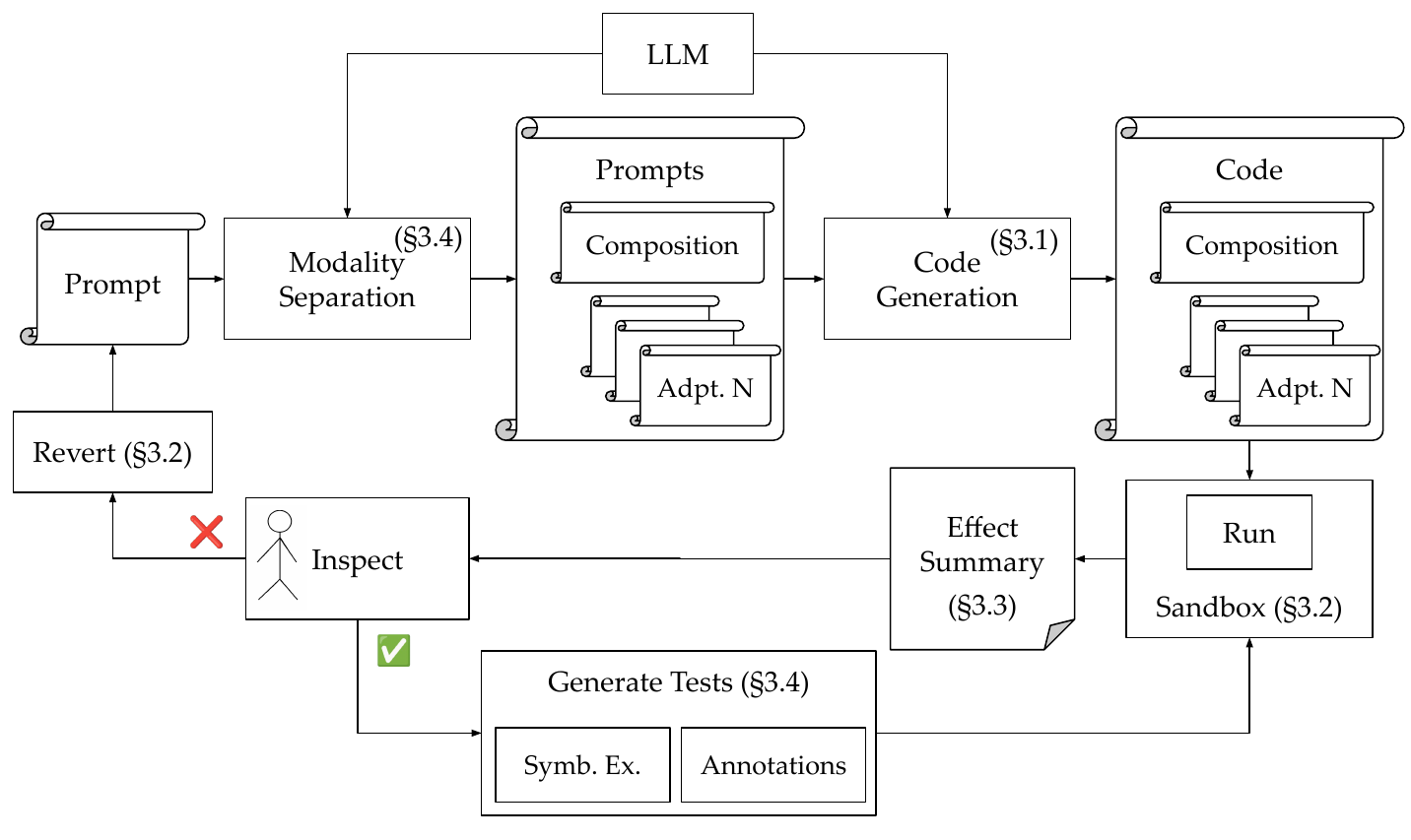}
    \caption{\sys architecture.
    }
    \label{fig:architecture}
\end{figure}

\section{\sys{}: A Natural Language Shell}
\label{sec:nash}

To address the challenges that shell users face, we propose to develop \sys{}.
\sys{} improves the \textbf{Develop}, \textbf{Inspect}, \textbf{Revert}, and \textbf{Generate Inputs} phases of shell use through the following components: generating code artifacts~\sx{s:code}, sandboxing with inverse overlays~\sx{s:sandbox}, effect summaries~\sx{s:effect_summaries}, and symbolic execution for test generation~\sx{s:test_generation}.
Although we present \sys{} as a single system, its components could be used independently outside of \sys{}'s overall architecture (e.g., \sys{}'s sandbox could be used for user-written code).

\subsection{Generating Code Artifacts}
\label{s:code}

A central question in designing \sys{} is how it should interact with the system on behalf of the user.
One could think of an LLM agent directly issuing system calls, taking a role similar to a graphical shell, or interacting with the system like a user would, at the level of mouse clicks or keystrokes~\cite{anthropiccomputeruse,googleduplex}.
The problem with these approaches is the \emph{lack of explainability and repeatability}.
If the LLM reacts incorrectly to the result of a syscall or mouse click, it is hard for the user to debug and correct what went wrong.
This is exacerbated by the fact that LLMs may produce \emph{different outputs} on different runs with the same input, due to inherent nondeterminism in state-of-the-art LLM serving techniques~\cite{chann2023nondeterminism, ivchenko2024nondeterminism, puigcerver2024softmoe}.

Instead, our approach in \sys{} is to have the LLM generate a \emph{code artifact} based on the user's prompt.
Some existing systems~\cite{gorillacli, engshell} take this approach, first generating code and then executing it.
But, \sys{} has an additional insight---the generated code is not only executable, but also \emph{analyzable}.
Specifically, \sys{} can draw on program analysis techniques to help users course correct if the shell does not behave as intended.
Furthermore, the code can be re-executed predictably even if the underlying LLM is nondeterministic, providing explainability and facilitating reusable script development.
If all else fails, a user with software experience can use the code as a starting point to build a correct solution.

\subsection{Sandboxing}
\label{s:sandbox}

\sys{} executes LLM-generated code in a \emph{sandbox} that provides two important pieces of functionality.
First, it limits \emph{at runtime} the damage that incorrect or malicious code may do.
Second, the sandbox gives users the ability to \emph{undo} an incorrect operation so that they can debug and try again.

We envision the ``undo'' functionality as a transaction-like extension of \texttt{history} functionality present in today's shells.
We maintain resources for each operation until the user decides to ``abort'' (i.e., undo) or ``commit'' (i.e., give up the ability to undo and release the resources for the operation).

\heading{Desiderata}
A good solution will have two properties.
First, the performance overhead of executing in the sandbox should be small, especially for the common case,
where the user does not undo operations.
Second, users should be able to delay ``committing'' or ``aborting'' for as long as possible.
This is important because LLMs distance users from precise execution logic, so users may not notice an issue until after many intervening operations have been performed.

\heading{Filesystem Effects}
\sys{}'s sandbox must isolate mutations to file system state.
One approach is to use \emph{overlay file systems}~\cite{pendry1995union}, which are used to virtualize the file system in OS container systems like LXC~\cite{lxc} and Docker~\cite{docker}.
The idea is to create a fresh overlay for each shell command, isolating its file system modifications to the overlay layer.
For operations meant to be ``read only'' (e.g., searching for a file should not modify the file system), the overlay layer can be dropped once execution completes.
This gives the generated code the ability to use temporary files and scratch space, while guaranteeing that all such side effects are removed.

An overlay-based approach can also support undoing commands meant to modify file system state.
The idea is to maintain the overlay layer until the user decides to commit or abort.
``Commit'' would mean copying the overlay contents to the base file system, and ``abort'' would mean to simply discard the overlay layer.

However, an overlay-based approach has important shortcomings.
First, file system modifications made by a shell command are confined to the overlay in which it executes, and are not visible to other applications running using the base file system without the overlay.
The modifications only become visible to other applications once the operation commits---but at that point, the user loses the ability to revert the changes.
This is not compatible with our desired approach of allowing the user to delay ``committing'' or ``aborting'' for as long as possible.
We would like the user to be able to perform an operation and use the file system as normal with the operation's effects, and undo it much later once an issue is discovered.
Second, overlays \emph{stack}.
After many uncommitted operations, the file system will have many layers, and each access will have to check them linearly to find the desired file.
Additionally, changes to ``special files'' (e.g., configuration in \texttt{/proc} or \texttt{/sys}) will not take effect with an overlay.

\heading{Inverse Overlays}
Instead, we propose a new solution that we call an \emph{inverse overlay}---the \emph{inverse} of an overlay file system.
Whereas an overlay filesystem stores the \emph{new} versions of all modified files in a layer and keeps the base filesystem \emph{unmodified}, an inverse overlay file system directly applies modifications to the underlying file system but stashes \emph{old} versions of modified files in a layer.
Each shell command would execute in its own \emph{inverse overlay} layer, and collectively the inverse overlay layers form an undo log~\cite[pp. 603]{anderson2014reliability} of sorts.
``Committing'' an operation discards its inverse overlay layer.
``Aborting'' an operation restores the contents of the inverse overlay layer to the base file system.

\sys{} lets users undo an operation independently of successor operations.
Furthermore, users can selectively restore certain files for an operation while keeping other changes intact.
Restoring a file to an older version can cause application-level inconsistencies if other operations took actions based on the file's newer version.
\sys{} leaves it to users to identify such inconsistencies and resolve them as they see fit.
In \secref{s:conflicts}, we discuss how \sys{} might help with this.

Unlike overlays, \emph{inverse overlay layers do not stack}, so there is no performance penalty to keeping a long history.
The only constraint is disk space; ``committing'' is like emptying a recycle bin, and can be automated (e.g., committing the oldest layers when disk space exceeds a threshold).
Thus, inverse overlays give users freedom to undo operations only as issues are discovered, even if those issues emerge long after the fact.

\heading{Other Effects}
Scripts may have effects that do not manifest in the file system, such as making external API calls or interacting with other system processes.
\sys{} takes the following approach for such effects: It prevents them, except if the user explicitly gives their permission.
To prevent such effects, \sys{} will run the sandboxed code in clean network, user, and pid namespaces.
If the user wishes to allow a subset of these effects for some commands in their script,
e.g., when running a data analytics task over a public dataset that needs to be downloaded~\cite[Fig. 2]{vasilakis2021pash}, \sys{} will allow them to selectively opt-out of namespaces.
Such control is coarse-grained (API calls are either completely allowed or not at all); we discuss how to provide more fine-grained control in \secref{s:api_sandbox}.

\subsection{Effect Summaries}
\label{s:effect_summaries}

\sys{} improves on the \textbf{Inspect} phase of shell use through a succinct effect summary language representing all changes in the system including additions and deletions of files or directories and modification of file data or metadata.
Since \sys runs all user commands and scripts in a sandbox, all command effects can be retrieved by summarizing the resulting inverse overlay layers. 
An effect summary localizes all effects and significantly improves the inspection phase since users do not need to manually determine where changes have happened, potentially missing any of them.
Having localized the effects, the user can then use LLM support to issue commands to better understand the effects, e.g., by inspecting and analyzing the contents of modified files and directories.

\begin{figure}[t]
    \centering
    \begin{minted}[frame=lines, framesep=2mm, fontsize=\footnotesize]{bash}
# Iterate over all files in the current directory
for file in *; do
  # Check if it's a regular file and not a .swp file
  if [[ -f "$file" && "${file##*.}" != "swp" ]]; then
    # If there is not a corresponding .swp file
    if [[ ! -f ".$(basename "$file").swp" ]]; then
      rm "$file"
    fi
  fi
done
    \end{minted}
    \caption{An example script that deletes all files in a directory that are not currently edited using Vim.}
    \label{fig:control-flow-script-example}
\end{figure}

\subsection{Test Generation}
\label{s:test_generation}

The sections so far form a robust solution for interactive use, but \emph{script development} (\secref{s:scripting}) has additional challenges since the script must generalize correctly to a wider set of inputs and environments.
Users are then tasked with determining useful inputs, and generating them in a reversible way.
\sys{} addresses this challenge by: (1) proposing a combination of symbolic execution and command annotations to explore the script input space, 
and (2) using its sandboxing mechanism to create throwaway testing environments.

\heading{Script Modalities}
Scripts often use external commands to perform their tasks.
A key challenge when exploring the relevant script input space is that commands can be arbitrary black-box binaries, so their code cannot be analyzed in an automated way to explore relevant inputs.
This leads to developers having to blindly generate random inputs and environments to test their commands, not adequately exploring their input space.
To address this challenge, \sys makes the following observation:
scripts contain two modalities, \emph{adapter tasks} and \emph{composition code}, each of which has very distinct development challenges.
Adapter tasks perform computation, using external components and low-level language features.
Composition code describes the control and data flow of the script across adapter tasks.
In \Cref{fig:control-flow-script-example}, the \ttt{rm} invocation and the \ttt{if} conditions are the adapter tasks, while the control flow surrounding them is the composition.

\heading{Code Generation}
\sys{} proposes to decouple code generation in two phases, adhering to the differences of the two modalities.
The first phase asks the LLM to split the original prompt into a workflow prompt and a set of task prompts.
The second phase then generates code from the prompts, a workflow composition from the workflow prompt, and adapter code from the task prompts.

\heading{Symbolic Execution}
In contrast to adapter code that uses external commands, composition code is fully analyzable.
\sys exploits that and proposes to develop a symbolic execution engine for composition code, akin to systems like KLEE~\cite{cadar2008klee}, by encoding it into an SMT formula.
The engine can then use an SMT solver to generate testing environments that better explore the script control flow, improving confidence of its correctness.

\heading{Command Annotations}
Even though external components are black boxes, parts of command behavior can be characterized and used for analyses.
This approach was pioneered by recent systems like PaSh~\cite{vasilakis2021pash} and POSH~\cite{posh} that develop annotation libraries for commands, describing properties like the command options and their inputs and outputs.
\sys can use such annotation libraries to better explore relevant input flags and options, as well as relevant command inputs.
As a fallback mechanism, \sys will provide random input and environment generation to test the validity of commands without annotations.

\section{Discussion}
\label{sec:discussion}

In this section, we identify open problems of LLM-powered shells and propose promising avenues for future research.

\subsection{Incorporating User Feedback}

During standard shell use, developers don't completely rewrite the script each time they revert, but they incrementally modify  it given the inspection results.
Such incremental user feedback is harder to incorporate when the code is generated through an LLM: a user might ask the LLM to slightly modify the code, but given its statistical nature it might return a more drastically rewriten script.
However, one way \sys{} can significantly improve the feedback process is that it can support multi-modal feedback from developers, not just in the form of prompts, but also in the form of input/output examples, which it could then add in a test set and always execute after generation (similarly to syntax guided synthesis~\cite{alur2013syntax}).
\subsection{Detecting ``Undo'' Conflicts}
\label{s:conflicts}

As discussed in \secref{s:sandbox}, undoing file operations may result in inconsistencies.
We can help the user identify certain inconsistencies by tracking file modifications and prompting the user in case of a conflict.
For example, suppose a user wants to undo an operation $x$, but a file modified by $x$ was modified by a subsequent operation $y$.
The system can detect the conflict by storing a version number and unique ID for each file in the base file system and inverse overlay, and comparing version numbers when restoring an old file version.
However application level conflicts are harder to detect; e.g., if an operation $x$ installs a package, and a subsequent operation $y$ installs a package that depends on it, then undoing only $x$ would leave the package manager in an inconsistent state.
Detecting these conflicts would require extensive application tracing, e.g., tracing all file reads, and application hints.

\subsection{External API Calls}
\label{s:api_sandbox}

Users may want to use \sys{} for tasks that access external state, such as data analytics over a public dataset~\cite[Fig. 2]{vasilakis2021pash}.
API calls, however, 
(1) cannot easily be sandboxed or undone, and (2) could externally leak private information.
Both of these are very challenging to address in a general and automated way.
We propose a semi-automatic approach that \emph{prompts the user} before making the API calls.

\heading{Identifying API Calls}
The first challenge is that identifying API calls is not possible through syscall tracing since they can be encrypted with TLS.
To address that, an LLM-powered shell should be limited to perform API calls only through a set of accepted tools like \texttt{curl} or the Python \texttt{requests} library.
This allows visibility to the call interface and arguments.

\heading{Securing Private Data}
LLM-generated code might mistakenly expose private user data via an API call, e.g., by passing the user's SSH private key to the request parameters.
\sys{}'s code artifact generation and command annotations enable a solution path for this: applying information flow control (IFC)~\cite{myers1999jflow, efstathopoulos2005asbestos, zeldovich2006histar, enck2010taintdroid} to indicate any local files from which the API call arguments were derived.

\heading{Describing API Calls to Users}
If prompted with raw HTTP requests, users might not understand the high-level semantics.
One solution to address this could be using data mining techniques~\cite{robillard2015recommending} to retrieve API documentation and compose it with the HTTP request, providing additional context to the user~\cite{head2015tutorons}.

\heading{Minimizing User-Facing Prompts}
Prompting the user once for each API call in a generated script might prove burdensome.
To address that we propose batching API prompts combining ideas of out-of-order execution~\cite{hs2023hotos} and external synchrony~\cite{nightingale2006externalsynchrony}.
After encountering the first API call, the runtime can \emph{skip its execution and the execution of any code that depends on its result}, and instead execute other code paths that do not depend on it.
The runtime only prompts the user when all codepaths are blocked by API calls.
For the data analysis workload from PaSh~\cite[Fig. 2]{vasilakis2021pash}, this technique would issue a single prompt for all four API calls.

\subsection{Toward Guardrail-Friendly APIs}

We speculate that, in the future, a significant fraction of API calls will be issued by LLM-based systems like \sys{}.
We discuss how APIs and RPC interfaces can be enhanced to enable undoing external calls, to support such a future.
We envision that the RPC stack can support these extensions, to ease the burden on developers.

\heading{Undo Stubs}
The first step is to extend API interfaces to support ``undo.''
RPC frameworks can provide both ``do'' and ``undo'' stubs at the server for each API call, requiring the developer to implement an ``undo'' routine.
Its exact behavior would depend on application semantics.
Some API calls may support undoing only within a time bound, while others (e.g., cloud APIs) may support undoing an action but not the payment, and still others may not support undoing at all.

\heading{Undo Contexts}
All API calls can return an \emph{undo context} with their response,
describing the constraints on undoing them (e.g., time limits) and information to initiate an ``undo.''
Undo contexts can be propagated through API call graphs similarly to Go \texttt{context}s~\cite{gocontexts}.
Specifically, when one service invokes others, it propagates undo contexts by combining the callee's undo contexts (e.g., choosing the earliest undo deadline), including constraints from its own execution, and returning the combined undo context to its client.
The RPC framework can provide syntax and tools to ease the use of undo contexts.

\section{Related Work}

\sys{} draws inspiration from prior work offering affordances for shell development, such as Autobash~\cite{autobash}, which can track causal dependencies during script development to help fix misconfigurations.
In addition, \sys{}'s sandboxing and proposed handling of external API calls bears similarities to Speculator~\cite{speculator} and External Synchrony~\cite{nightingale2006externalsynchrony}.
Both works propose ideas for managing local and external effects through the use of checkpointing.
\sys{} builds on all these ideas with its effect summaries and sandboxing, and provides additional script development affordances, particularly targeting LLM-powered shells, such as input generation.

There have been previous works developing sandboxing mechanisms for shell commands, such as try~\cite{hs2023hotos}, GoEX~\cite{patil2024goex}, MBOX~\cite{mbox}, and other syscall interposition systems~\cite{goldberg1996janus,jain2000user,garfinkel2004ostia}. 
These could be used as alternatives for \sys{}'s sandbox, offering some benefits like more flexible policies in the case of syscall tracing. 
However, \sys{} still makes key tradeoffs that improve sandboxing performance for its use case of LLM-powered shell development, improving some overheads of these alternative mechanisms:
MBOX interposes on all system calls, GoEX uses a git-based sandbox recursively scanning the file system to find changes, and try uses OverlayFS, which stacks~(\secref{s:sandbox}).
For API calls, GoEX proposes dry runs and an LLM-populated \texttt{Reversion Set}; these are complementary to the techniques proposed in our paper.

Finally, automated program generation has been a longstanding goal, with traditional program synthesis~\cite{alur2013syntax} techniques being widely used for it.
These techniques are not applicable in the context of the shell because they either require a specification of the program behavior or extensive input-output examples that are hard to construct for shell scripts that have complex effects on the system state. 
More recently, the emergence of LLMs has led to the development of LLM-powered shells by industry and open-source communities (e.g., Github CLI Copilot~\cite{gh-copilot-cli}, Warp~\cite{warp}, and Engshell~\cite{engshell}).
These shells offer significantly less support to the user than \sys{}, primarily asking them for permission before running every command.

\begin{acks}

We would like to thank Shishir G. Patil for formative conversations early in this project.
We would also like to thank all the people who work on the PaSh project for discussions that have helped shape our understanding of scripting, as well as for their work on try.
These include but are not limited to: Nikos Vasilakis, Michael Greenberg, Di Jin, Georgios Liargkovas, and Ezri Zhu.

\end{acks}

\bibliographystyle{ACM-Reference-Format}
\bibliography{nash}

\end{document}